\documentclass[12pt,preprint]{aastex}
\shorttitle{The helium abundance in Tol 1214--277 and Tol 65}
\shortauthors{Izotov et al.}
\begin{document}

\title{The $^4$He abundance in the metal-deficient blue compact dwarf
galaxies Tol 1214--277 and Tol 65\footnote{The 
observations reported here were obtained at the W. M. Keck Observatory, which
is operated as a scientific partnership among the California Institute of
Technology, the University of California, and the National Aeronautics and
Space Administration. The Observatory was made possible by the generous
financial support of the W. M. Keck Foundation.}}

\author{Yuri I. Izotov\footnote{Visiting astronomer, National Optical
Astronomy Observatory, which is operated by the Association of Universities for
Research in Astronomy, Inc. (AURA) under cooperative agreement with the
National Science Foundation (NSF).}}
\affil{Main Astronomical Observatory, National Academy of Sciences of Ukraine, 03680, Kyiv, Ukraine}
\email{izotov@mao.kiev.ua}

\author{Frederic H. Chaffee}
\affil{W. M. Keck Observatory, 65-1120 Mamalahoa Hwy., Kamuela, HI 96743, USA}
\email{fchaffee@keck.hawaii.edu}

\and

\author{Richard F. Green}
\affil{National Optical Astronomy Observatory, Tucson, AZ 85726, USA}
\email{rgreen@noao.edu}

%

\begin{abstract}
We present high-quality Keck telescope spectroscopic observations of the two
metal-deficient blue compact dwarf (BCD) galaxies Tol 1214--277 and Tol 65.
These data are used to derive the heavy-element and helium abundances. We find
that the 
oxygen abundances in Tol 1214--277 and Tol 65 are the same, 12 + log O/H 
= 7.54 $\pm$ 0.01, or $Z_\odot$/24, despite the different ionization conditions
in these galaxies. The nitrogen-to-oxygen abundance ratio in both galaxies is 
log N/O = --1.64$\pm$0.02 and lies in the narrow range found for the other 
most metal-deficient BCDs. We use the five strongest He {\sc i}  emission lines
$\lambda$$\lambda$3889, 4471, 5876, 6678 and 7065, to correct self-consistently
their intensities 
for collisional and fluorescent enhancement mechanisms and to derive the
$^4$He abundance. Underlying stellar absorption is found to be important 
for the He {\sc i} $\lambda$4471 emission line in both galaxies, being larger 
in Tol 65. The weighted $^4$He mass fractions in Tol 1214--277 and
Tol 65 are respectively $Y$ = 0.2458 $\pm$ 0.0039 and 0.2410 $\pm$ 0.0050 when 
the three He {\sc i} emission lines, $\lambda$$\lambda$4471, 5876 and 6678 \AA,
are used, and are, respectively, 0.2466 $\pm$ 0.0043 and
0.2463 $\pm$ 0.0057 when the He {\sc i} 4471 \AA\ emission line is excluded.
These values are in very good agreement with recent measurements of the $^4$He
mass fraction in others of the most metal-deficient BCDs by Izotov and 
coworkers. We find that the combined effect of the systematic 
uncertainties due to the underlying He {\sc i} stellar absorption lines, 
ionization and temperature structure of the H {\sc ii} region and collisional 
excitation of the hydrogen emission lines is likely small, not exceeding 
$\sim$ 2\% (the error is 2$\sigma$). 
Our results support the validity of the standard big bang model of
nucleosynthesis.
\end{abstract}


\section {Introduction}

    In the standard big bang model of nucleosynthesis (SBBN), the
light isotopes D, $^3$He, $^4$He and $^7$Li were produced by nuclear reactions
a few minutes after the birth of the Universe. Given the number of
light neutrino species $N_\nu$ = 3 and the neutron lifetime, the abundances of
these light elements depend on one cosmological parameter only, the
baryon-to-photon ratio $\eta$, which in turn is directly related to the
density of ordinary baryonic matter $\Omega_b$.

  The ideal objects for determination of the primordial helium abundance 
are blue compact dwarf (BCD) galaxies.
These dwarf systems are the least chemically
evolved galaxies known, so they contain very little helium manufactured by
stars after the big bang. Because the big bang production
of $^4$He is relatively insensitive to the density of matter, the primordial
abundance of $^4$He must be determined  to very high precision (better 
than a few percent relative accuracy) in order to put useful constraints on 
$\Omega_{\rm b}$. 
This precision can be achieved with very high signal-to-noise ratio optical 
spectra of BCDs. These BCDs are undergoing intense bursts of star formation, 
giving birth to high excitation supergiant H {\sc ii} regions, which allow an 
accurate determination of the helium abundance in the ionized gas through the 
BCD's emission-line spectrum.
The primordial helium mass fraction $Y_{\rm p}$ of $^4$He is usually derived
by extrapolating the $Y$ -- O/H and $Y$ -- N/H correlations to O/H =
N/H = 0, as proposed originally by Peimbert \& Torres-Peimbert (1974, 1976)
and Pagel, Terlevich \& Melnick (1986). Many attempts at determining
$Y_{\rm p}$ have been made, using these correlations on
various samples of dwarf irregulars and BCDs (e.g., Pagel et al. 1992; 
Izotov, Thuan \& Lipovetsky 1994, 1997, hereafter ITL94 and ITL97; 
Olive, Steigman \& Skillman 1997; Izotov \& Thuan 1998; Pagel 2000). Another
way to infer the primordial $^4$He abundance is to measure the helium 
abundance in the most metal-deficient BCDs, which is very close to the
primordial value (e.g., Izotov et al. 1999).

    Recent $Y_{\rm p}$ determinations made by Izotov \& Thuan (1998) and
Izotov et al. (1999) resulted in a very narrow range of
$Y_{\rm p}$ $\sim$ 0.244 -- 0.245. They used high signal-to-noise 
spectroscopic observations of BCDs reduced in a homogeneous way.
A self-consistent method was applied to correct He {\sc i} emission
line intensities for the collisional and fluorescent enhancement mechanisms
which lead to deviations from the values predicted
by recombination theory. The use of several He {\sc i}
lines allows one to discriminate between collisional and fluorescent
enhancements which change the line intensities in different ways. It also
allows one to estimate the importance of underlying stellar absorption
in each galaxy and to improve the precision of the $Y_{\rm p}$
determination through the use of several lines. The details of this
approach are discussed by ITL97 and Izotov \& Thuan (1998).

    Our present study is the continuation of helium abundance determinations
in the lowest-metallicity BCDs based on spectroscopic observations with
the 10m Keck telescope. Earlier such studies have been done for the BCDs 
SBS 0335--052 ($Z_\odot$/40, Izotov et al. 1999) and SBS 0940+544 
($Z_\odot$/27, Guseva et al. 2001). We focus here on two southern BCDs, 
Tol 1214--277 and Tol 65 $\equiv$ Tol 1223--359, the very low 
metallicity of which ($\sim$ $Z_\odot$/25) has been established by earlier 
spectroscopic work (Kunth \& Sargent 1983; Campbell et al. 1986; Pagel et al. 
1992; Masegosa, Moles \& Campos-Aguilar 1994; Fricke et al. 2001). The 
motivation of our work is as follows. First, these galaxies are
relatively bright targets for a large telescope which allows the derivation of 
the $^4$He abundance with great precision. Second, Tol 1214--277
and Tol 65 are the second and third lowest-metallicity BCDs in the Pagel et
al. (1992) sample (after I Zw 18), for which those authors derived very low 
$^4$He mass fractions $Y$ = 0.233 and 0.231 respectively. Those values are in
disagreement with the significantly larger $Y$'s derived in later studies 
of other very low metallicity BCDs. Third, recently several papers have
appeared (e.g., Ballantyne, Ferland \& Martin 2000; Viegas, Gruenwald \&
Viegas 2000; Sauer \& Jedamzik 2001; Peimbert, Peimbert \& Luridiana 2001;
Stasi\'nska \& Izotov 2001) where systematic effects on the $^4$He
abundance determination are discussed. Here we estimate the systematic 
uncertainties for several of the best-observed low-metallicity BCDs,
including Tol 1214--277 and Tol 65.

In Sect. 2 we describe the observations and data reduction. Heavy element
abundances are derived in Sect. 3. The results of the $^4$He abundance
determination are discussed in Sect. 4. We discuss the systematic uncertainties
in $^4$He abundance determinations in Sect. 5. The summary is in Sect. 6.

%
\section{Observations and data reduction}

\begin{figure}
\epsscale{1.11}
\figurenum{1}
\plottwo{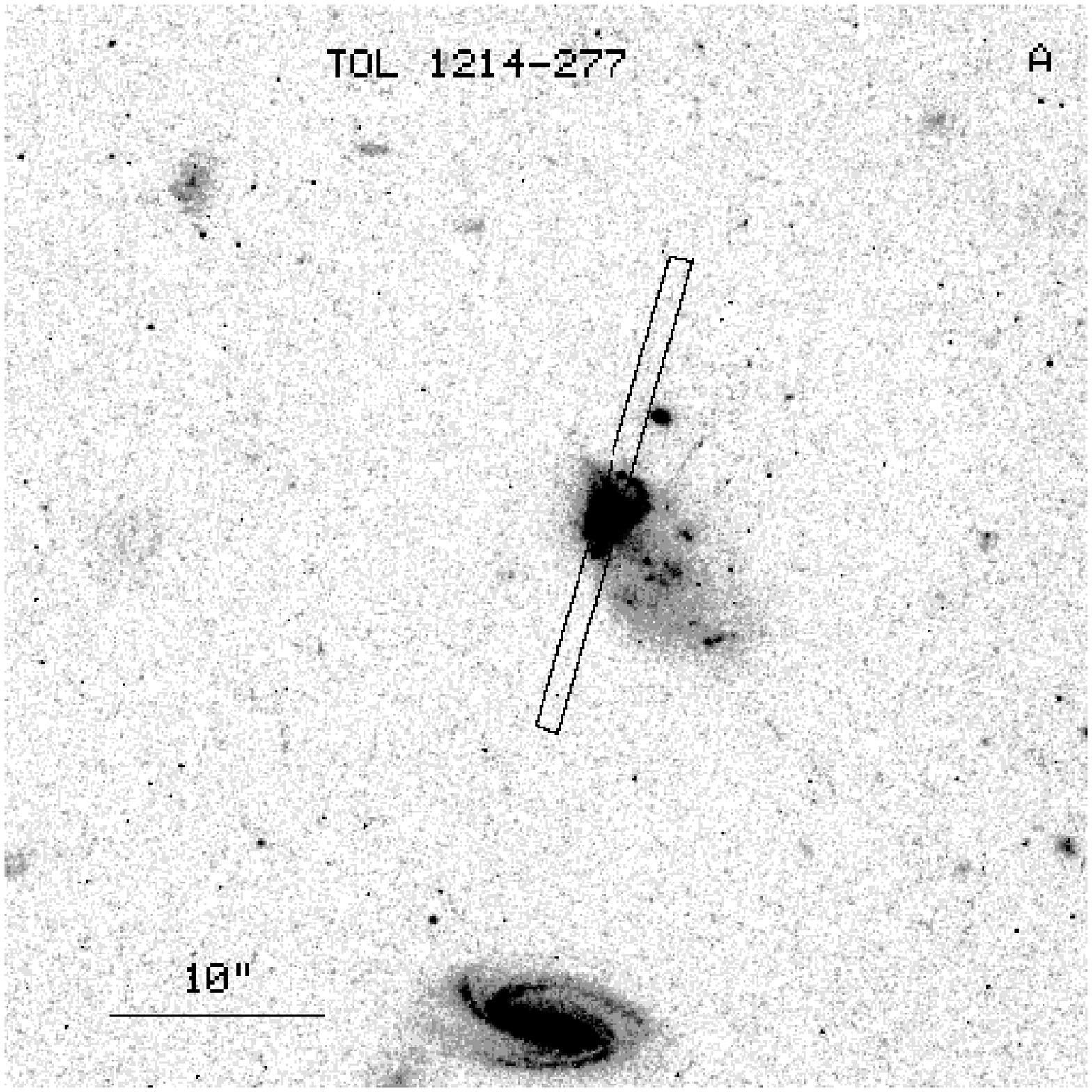}{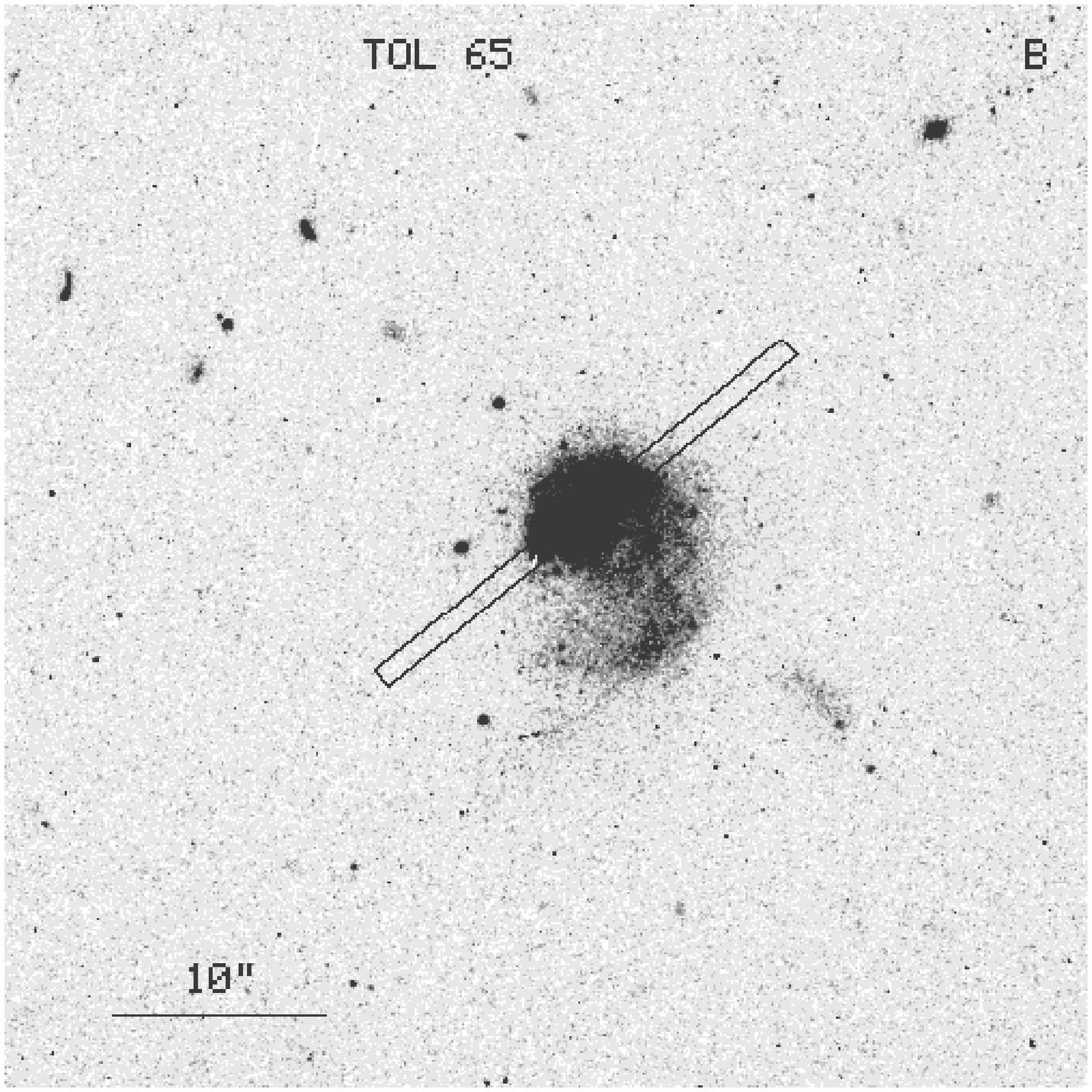}
\caption{\label{Fig1} The archival {\sl Hubble Space Telescope} $V$ images
of Tol 1214--277 (a) and Tol 65 (b) with the orientation of the slit
overplotted.}
\end{figure}

 The Keck II spectroscopic observations of Tol 1214--277 and Tol 65 were 
carried out on January 9, 2000, with  the low-resolution imaging spectrograph 
(LRIS) (Oke et al. 1995), using the 300 groove mm$^{-1}$ grating, which 
provides a dispersion 2.52 \AA\ pixel$^{-1}$ and a spectral resolution of 
about 8 \AA\ in first order. The slit was 1\arcsec$\times$180\arcsec, centered 
on the brightest central regions and oriented with 
a position angle P.A. = --19$^{\circ}$ for Tol 1214--277 and --51$^{\circ}$
for Tol 65 (Fig. \ref{Fig1}). No binning along the spatial axis has been done, 
yielding a spatial sampling of 0\farcs2 pixel$^{-1}$. The total exposure time 
was 45 min for each galaxy, broken into three 15-min exposures. 
All exposures were taken at airmasses of
1.7 and 1.8 for Tol 1214--277 and Tol 65, respectively. The seeing was 
0\farcs9. No blocking filter was used. Therefore some second-order
contamination is present in the red parts of spectra at wavelengths 
$>$ 7000\AA.

The spectrophotometric standard stars Feige 34 and HZ 44 were observed for 
flux calibration. Spectra of a Hg-Ne-Ar comparison lamp were obtained before 
and after each observation to provide the wavelength calibration.

\begin{figure}
\epsscale{1.0}
\figurenum{2}
\plotone{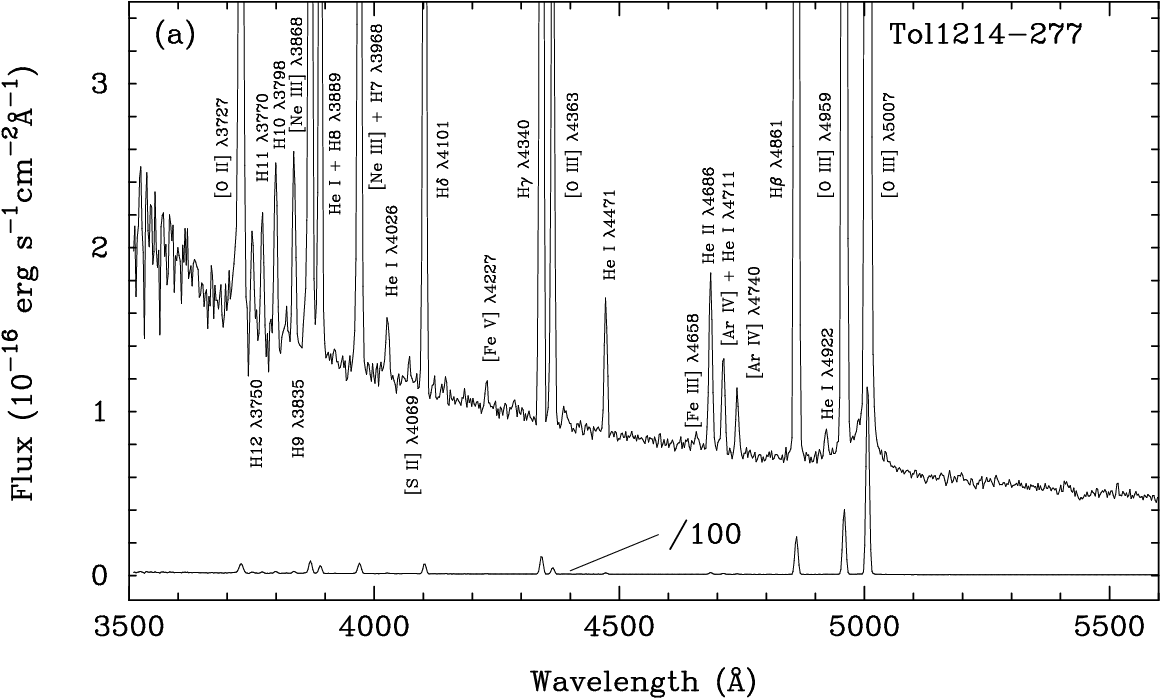}
\plotone{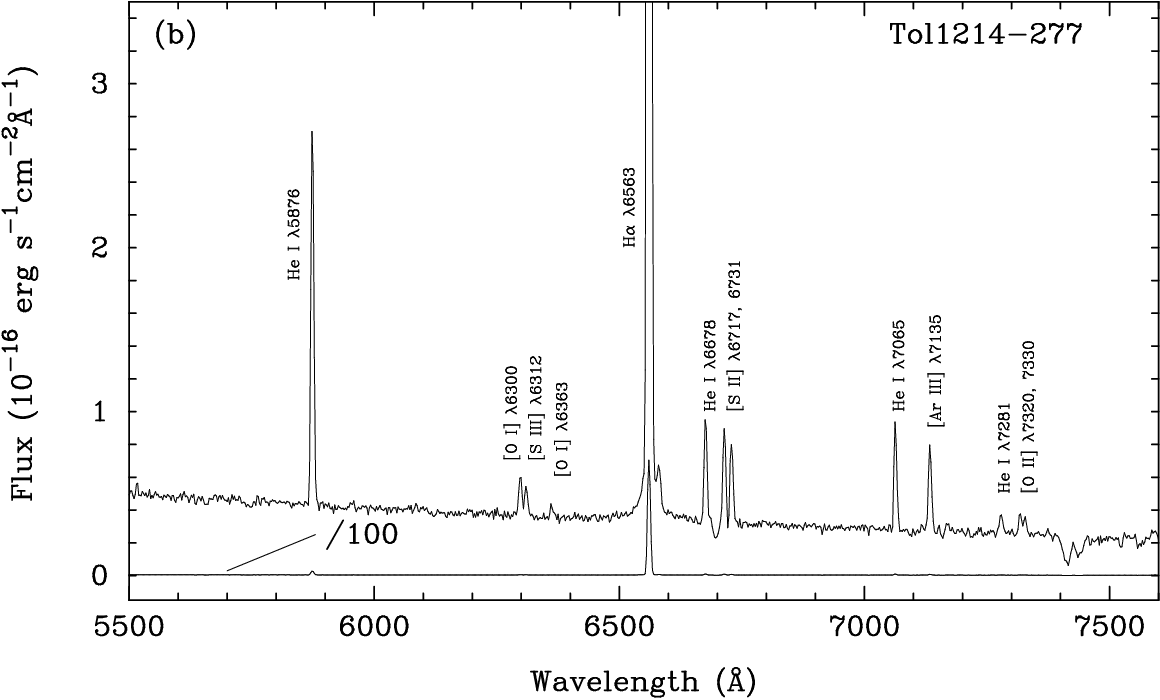}
\caption{\label{Fig2} The spectrum of Tol 1214--277 in aperture
1\arcsec\ $\times$ 6\farcs8.
The lower spectrum is the observed spectrum downscaled by a factor of 
100.}
\end{figure}

\begin{figure}
\epsscale{1.0}
\figurenum{3}
\plotone{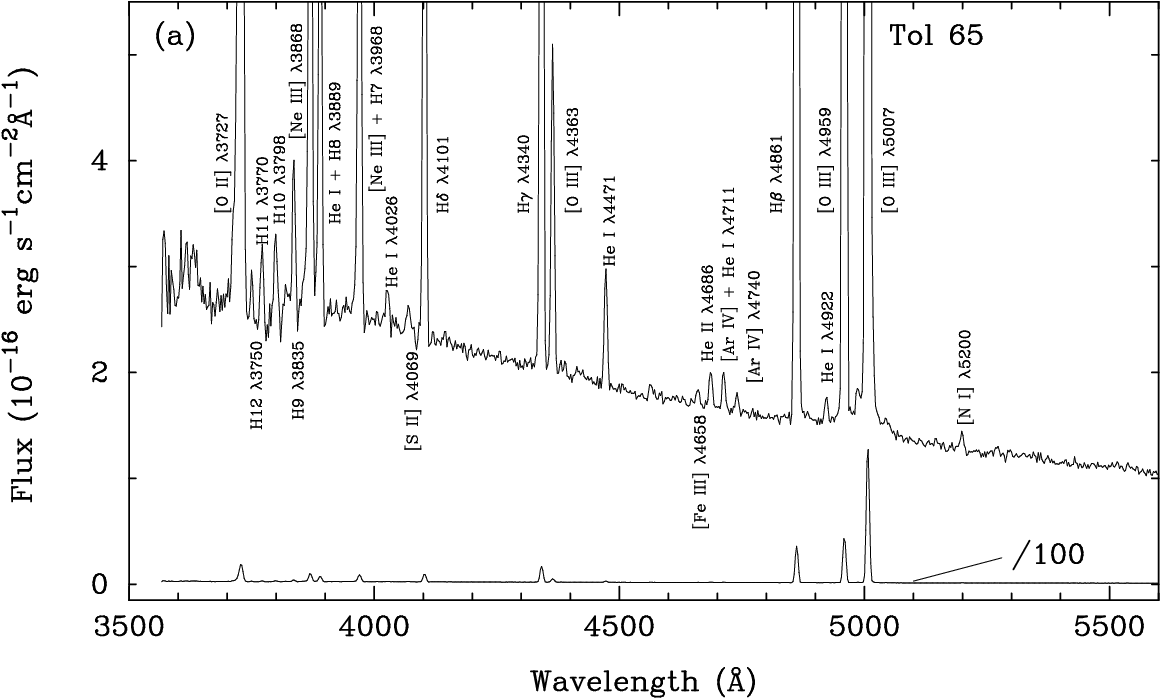}
\plotone{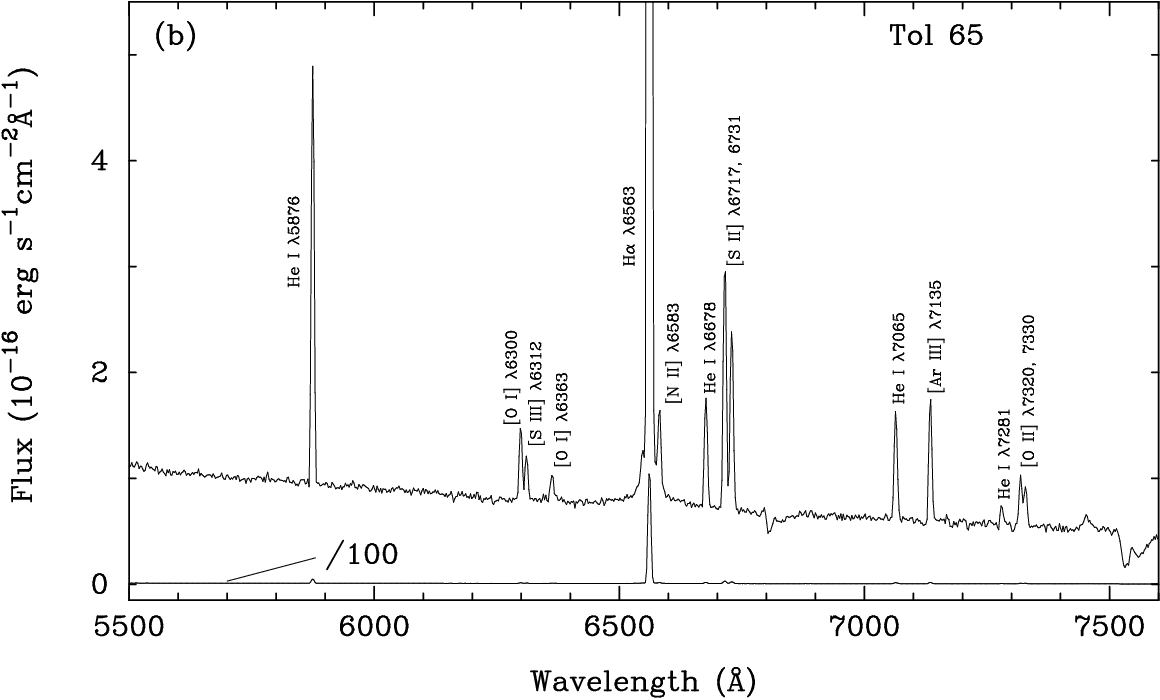}
\caption{\label{Fig3} The spectrum of Tol 65 in aperture
1\arcsec\ $\times$ 8\farcs4.
The lower spectrum is the observed spectrum downscaled by a factor of 
100.}
\end{figure}

Data reduction of the observations was carried out at the Main
Astronomical Observatory of Ukraine using the IRAF 
software package.\footnote{IRAF: the Image Reduction
and Analysis Facility is distributed by the National Optical Astronomy 
Observatory.} 
This reduction included bias subtraction, cosmic-ray removal
and flat-field correction using exposures of a quartz incandescent
lamp. After wavelength calibration, night-sky background subtraction, 
and correction for atmospheric extinction, each frame was
calibrated to absolute fluxes. Because both BCDs were observed at
large airmasses, the atmospheric differential refraction can be important
(Filippenko 1982). This effect is smaller for Tol 1214--277 because of its
higher declination and slit position angle close to the
parallactic one. To minimize the effect of the atmospheric differential 
refraction we extracted one-dimensional spectra from large apertures 
1\arcsec\ $\times$ 6\farcs8 for Tol 1214--277 and
1\arcsec\ $\times$ 8\farcs4 for Tol 65. The spectra are shown in
Figs. \ref{Fig2} and \ref{Fig3}. 
They are dominated by very strong emission lines. 
Remarkable spectral features in Tol 1214--277 are the strong nebular 
He {\sc ii} $\lambda$4686 and [Fe {\sc v}] $\lambda$4227 emission lines
suggesting a very hard stellar radiation field in the BCD (Fig. \ref{Fig2}). 
The latter line was detected first by Fricke et al. (2001) in the VLT spectrum
of Tol 1214--277. This line is absent in the spectrum of Tol 65 where 
He {\sc ii} $\lambda$4686 is also weaker, implying that the radiation in
this galaxy is softer.

  The fluxes of the nebular lines have been measured by fitting Gaussians to 
the line profiles. The errors in the line fluxes include the errors
in placement of the continuum and those in the Gaussian fitting. We also take
into account the errors introduced by uncertainties in the spectral energy
distributions of the standard stars. Standard star flux deviations for both 
Feige 34 and HZ 44 are taken to be 1\% (Oke 1990; Bohlin 1996). These 
1$\sigma$ errors
have been propagated in calculations of the electron temperature,
electron number density and elemental abundances.
The observed and extinction-corrected emission line fluxes and 
observed equivalent widths are shown in Table \ref{Tab1} together with the
extinction coefficient $C$(H$\beta$), derived from the decrement 
of hydrogen emission lines which includes both Milky Way and target galaxy
extinction, the absolute flux $F$(H$\beta$) of the
H$\beta$ emission line and average equivalent width $EW$(abs) of the hydrogen 
absorption lines. The errors of the fluxes and equivalent widths of 
the emission lines introduced by second-order contamination are relatively
small. We estimate this effect by measuring the fluxes of the [O {\sc ii}]
$\lambda$3727 and [Ne {\sc iii}] $\lambda$3868 emission lines in the second 
order spectra of both galaxies and find that they are respectively 
$\sim$ 1.5\% and $\sim$ 3 -- 4\% of those in the first order spectra. 
This implies that the effect of the 
second-order contamination is less than 1\% at wavelengths shorter 7500\AA,
smaller than the flux errors of the weak lines seen at 7000 -- 7300\AA.
Therefore, we do not take into account this effect.

%
\section{Heavy element abundances}

To derive heavy element abundances, we have followed the procedure detailed in 
ITL94 and ITL97. We adopted a two-zone photoionized H {\sc ii} region model 
(Stasi\'nska 1990) including a high-ionization zone with temperature 
$T_{\rm e}$(O {\sc iii}), and a low-ionization zone with temperature 
$T_{\rm e}$(O {\sc ii}). We have determined $T_{\rm e}$(O {\sc iii}) from the 
[O {\sc iii}]$\lambda$4363/($\lambda$4959+$\lambda$5007) ratio 
using a five-level atom model. That temperature is used for the 
derivation of the O$^{+2}$, Ne$^{+2}$ and Ar$^{+3}$ ionic 
abundances. 
To derive $T_{\rm e}$(O {\sc ii}), we have utilized the relation between
$T_{\rm e}$(O {\sc ii}) and $T_{\rm e}$(O {\sc iii}) (ITL94), based on a fit 
to the photoionization models of Stasi\'nska (1990). The temperature 
$T_{\rm e}$(O {\sc ii}) is used to derive the O$^+$, N$^+$, S$^+$ 
and Fe$^+$ ionic abundances. For Ar$^{+2}$ and S$^{+2}$ we have adopted an 
electron temperature intermediate between $T_{\rm e}$(O {\sc iii}) and 
$T_{\rm e}$(O {\sc ii}) following the prescriptions of Garnett (1992). 
The electron number density $N_{\rm e}$(S {\sc ii}) (Table \ref{Tab2}) is 
derived from the [S {\sc ii}] $\lambda$6717/$\lambda$6731 ratio. 
We point out that the flux of the [S {\sc ii}] $\lambda$6717 emission line
in Tol 1214--277 spectrum is significantly reduced due to coincidence
with a night sky absorption line. This artificially low intensity results in 
a significant overestimate of $N_{\rm e}$(S {\sc ii}). However, the heavy
element abundances in low-density H {\sc ii} regions do not depend on 
$N_{\rm e}$. Therefore, the uncertainties in $N_{\rm e}$(S {\sc ii})
(Table \ref{Tab2}) do not contribute significantly to the error
budget of the heavy element abundances.

The oxygen abundance is derived as
\begin{equation}
{\rm \frac{O}{H} = \frac{O^+}{H^+} + \frac{O^{+2}}{H^+} + \frac{O^{+3}}{H^+}},
\label{eq:O} 
\end{equation}
where the O$^{+3}$ abundance is derived from the relation
\begin{equation}
{\rm \frac{O^{+3}}{O^+ + O^{+2}} = \frac{He^{+2}}{He^+}}.
\end{equation}

Total abundances of other heavy elements were computed after correction 
for unseen stages
of ionization as described in ITL94 and Thuan, Izotov \& Lipovetsky (1995).

The abundances of oxygen and other heavy elements obtained in this study for 
both galaxies are in general agreement with previous studies.
Our value for the oxygen abundance in 
Tol 1214--277, 12 + log (O/H) = 7.54 $\pm$ 0.01, compares with values of 
7.54$\pm$0.04 (Campbell et al. 1986), 7.59$\pm$0.05 (Pagel et al. 1992), 
7.57$\pm$0.01 (Masegosa et al. 1994), 7.58$^{+0.025}_{-0.026}$ (Kobulnicky \&
Skillman 1996), 7.52$\pm$0.01 (Fricke et al. 2001). In Tol 65 we derive an 
oxygen abundance 12 + log (O/H) = 7.54 $\pm$ 0.01 which is in
agreement with 7.53$\pm$0.05 (Kunth \& Sargent 1983), 7.42$\pm$0.07 
(Campbell et al. 1986), 7.59$\pm$0.05 (Pagel et al. 1992), 7.40 -- 7.54 
(Masegosa et al. 1994), 7.56$^{+0.032}_{-0.033}$ (Kobulnicky \& Skillman 1996).

    The spectral resolution of the Tol 1214--277 spectrum is not enough to
measure the flux of the weak [N {\sc ii}] $\lambda$6583 which is
contaminated by the H$\alpha$ emission line. Therefore, to derive the nitrogen
abundance we use the 
[N {\sc ii}]/H$\alpha$ flux ratio obtained by Pagel et al. (1992) from a
higher resolution spectrum. We
obtain log N/O = --1.64$\pm$0.02. This value is lower than previous values
of --1.46$\pm$0.06 (Pagel et al. 1992) and 
--1.45$^{+0.080}_{-0.098}$ (Kobulnicky \& Skillman 1996).
The difference in N/O comes mainly from the differing fluxes of 
[O {\sc ii}] $\lambda$3727 emission line. This line is used for 
determination of the ionization correction factor for nitrogen and it varies
from $I$($\lambda$3727)/$I$(H$\beta$) = 0.23 (Pagel et al. 1992) to
0.34 in this paper, the difference being 0.17 dex.
Fricke et al. (2001) derived log N/O = --1.50$\pm$0.02 from
VLT observations. However, no flux calibration was available for those
observations and hence Fricke et al. (2001) used earlier spectroscopic 
observations of Tol 1214--277 with the KPNO 2.1m telescope to calibrate their 
VLT spectrum. The [N {\sc ii}] 
$\lambda$6583 emission line is stronger in the spectrum of Tol 65 and we
derive log N/O = --1.64 $\pm$ 0.02. All other log N/O values for Tol 65 are
derived from a single observation by Kunth \& Sargent (1983) who obtained
log N/O = --1.75 $\pm$ 0.07, while Pagel et al. (1992) quote --1.81$\pm$0.15,
and Kobulnicky \& Skillman (1996) quote --1.79$^{+0.15}_{-0.23}$.
Our value of log N/O in Tol 65 obtained from the high signal-to-noise ratio
spectrum is the same as that derived for Tol 1214--277
and is very close to the mean value of --1.60, derived for the most 
metal-deficient galaxies with 12 + log O/H $<$ 7.6 (Thuan et al. 1995; 
Izotov \& Thuan 1999), further supporting the very low dispersion of the 
N/O ratio in those galaxies. Such a constant N/O abundance ratio
in the lowest-metallicity BCDs favors primary production of nitrogen in
massive stars (Izotov \& Thuan 1999) and may have important implications
for analysis of the abundance patterns in high-redshift damped
Ly$\alpha$ systems (Izotov, Schaerer \& Charbonnel 2001).

   The element-to-oxygen abundance ratios in Tol 1214--277 and Tol 65 
for $\alpha$-product elements
are the same within the errors (Table \ref{Tab2}) and they are close to the
mean values derived for low-metallicity BCDs (Izotov \& Thuan 1999). The 
exception is iron. While the Fe/O abundance ratio in Tol 65, though slightly
lower, is consistent with the mean value for BCDs (Izotov \& Thuan 1999), the
Fe/O abundance ratio in Tol 1214--277 is 2$\sigma$ larger. A similar Fe/O 
abundance 
ratio was found by Izotov et al. (1997, 1999) for the BCD SBS 0335--052 and
was interpreted to be a result of the contamination of the [Fe {\sc iii}] 
$\lambda$4658 emission line by stellar or nebular C {\sc iv} $\lambda$4658
line emission. This interpretation seems to be likely, because both
SBS 0335--052 and Tol 1214--277 are galaxies with the strongest nebular
He {\sc ii} $\lambda$4686 emission lines, implying very hard ionizing
radiation. Only in those two BCDs was the [Fe {\sc v}] $\lambda$4227 emission
line detected, again supporting the presence of hard radiation
(Fricke et al. 2001). Our derived abundance of O$^{+3}$ in Tol 1214--277 is 
5.5\% of the total oxygen abundance (Table \ref{Tab2}). Hence we expect that a 
significant amount of carbon in the H {\sc ii} region is present in the
form of C$^{3+}$, 
implying the presence of C {\sc iv} $\lambda$4658 line emission.

%
\section{Helium abundance}
%

He {\sc i} emission-line fluxes are converted to singly ionized helium 
abundances $y^+$ $\equiv$ He$^+$/H$^+$ using theoretical He {\sc i} 
recombination line emissivities by Smits (1996). However, collisional and 
fluorescent enhancements can cause the observed He {\sc i} fluxes to deviate 
from recombination values. In order to correct for these effects, 
we have adopted the following procedure,  
discussed in more detail in ITL94 and ITL97. We 
evaluate the electron number density $N_{\rm e}$(He {\sc ii}) and the 
optical depth $\tau$($\lambda$3889) in the He {\sc i} $\lambda$3889 line in a 
self-consistent way, so that the He {\sc i} $\lambda$$\lambda$3889/5876, 
4471/5876, 6678/5876 
and 7065/5876 line ratios have their recombination values, after correction 
for collisional and fluorescent enhancement.  Corrections are determined using 
the formulae by Kingdon \& Ferland (1995) for collisional enhancement and the 
Izotov \& Thuan (1998) fits to Robbins (1968) calculations for fluorescent 
enhancement. The He {\sc i} $\lambda$3889 and 7065 
lines play an important role because they are particularly sensitive to both 
optical depth and electron number density. Since the
He {\sc i} $\lambda$3889 line is blended with the H8 $\lambda$3889 line, 
we have subtracted the 
latter, assuming its intensity to be equal to 0.106 $I$(H$\beta$) (Aller 1984),
after correction for interstellar extinction and underlying stellar 
absorption in hydrogen lines. The singly ionized helium abundance $y^+$ 
and $^4$He mass fraction $Y$ is obtained for each of the three He {\sc i} lines
$\lambda$$\lambda$ 4471, 5876 and 6678. 
We then derive the weighted mean $y^+$ of these three determinations, the
weight of each line being determined by its intensity. However,
this weighted mean value may be underestimated due to the lower
value of $y^+$(4471) resulting from underlying stellar absorption. 
Therefore, in subsequent discussions
we also use the weighted mean values of $Y$ derived from the
intensities of only two lines, He {\sc i} $\lambda$5876 and $\lambda$6678 \AA.

   Additionally, we have added to $y^+$ the abundance of doubly ionized helium 
$y^{+2}$ which is derived from the He {\sc ii} $\lambda$4686 \AA\ emission line flux. 
Finally the helium mass fraction is calculated as
\begin{equation}
Y=\frac{4y[1-20({\rm O/H})]}{1+4y},                     \label{eq:Y}
\end{equation}
where $y$ = $y^+$ + $y^{+2}$ is the number density of helium relative to 
hydrogen (Pagel et al. 1992).

   The results of the $^4$He abundance determination are presented in Table
\ref{Tab3}, where we show the adopted electron temperature $T_{\rm e}$ and the
derived electron number density $N_{\rm e}$ in the He$^+$ zone, the optical
depth in the He {\sc i} $\lambda$3889 emission line, ionic abundances $y^+$ 
and $y^{+2}$, total abundances $y$ and helium mass fractions $Y$ derived for
each line and the two weighted means. 
The errors in $T_{\rm e}$(O {\sc iii}) and $N_{\rm e}$ (He {\sc ii}) are
propagated in calculations of the helium mass fractions $Y$.
The helium mass fraction in Tol 1214--277 is slightly lower for the He {\sc i} 
$\lambda$4471 emission
line, which is most subject to underlying stellar absorption. Similarly,
systematically lower $Y$ from the He {\sc i} $\lambda$4471 emission line was
derived earlier in two BCDs, SBS 0335--052 ($Z_\odot$/40, Izotov et al. 1999)
and SBS 0940+544 ($Z_\odot$/27, Guseva et al. 2001), observed with Keck.
The contribution of the doubly ionized He in Tol 1214--277 is significant and 
amounts to 6\% of the total He abundance. Earlier, Fricke et al. (2001) 
arrived at the same conclusion for this BCD. The helium mass fraction in 
Tol 1214--277 is $Y$ = 0.2458 $\pm$ 0.0039 when all three lines are used, 
and 0.2466 $\pm$ 0.0043 when the He {\sc i} $\lambda$4471 emission line is 
excluded. The effect of the underlying stellar absorption is more significant
in Tol 65. The helium mass fraction $Y$ derived from the He {\sc i} 
$\lambda$4471 emission line is $\sim$ 10\% lower than that derived from other
lines. The contribution of doubly ionized helium in Tol 65 is smaller than that
in Tol 1214--277 and amounts to $\sim$ 1.4\% of the total helium abundance.
The helium mass fraction in Tol 65 is $Y$ = 0.2410 $\pm$ 0.0050 when all three 
lines are used, and 0.2463 $\pm$ 0.0057 when the He {\sc i} $\lambda$4471 
emission line is excluded. Because of the effect of underlying absorption in 
the He {\sc i} $\lambda$4471 emission line, we finally adopt for Tol 1214--277 
and Tol 65 respectively $Y$ = 0.2466 $\pm$ 0.0043 and 0.2463 $\pm$ 0.0057.
These values are very similar to 0.2463 $\pm$ 0.0015 in SBS 0335--052 
(Izotov et al. 1999) and 0.2468 $\pm$ 0.0034 in SBS 0940+544 (Guseva et al. 
2001) derived from the analysis of high signal-to-noise ratio Keck spectra.

\section{Systematic effects}

 The similarity of $Y$ in the lowest-metallicity BCDs suggests that statistical
errors in the helium abundance determination are small. However, some 
systematic effects may change the value of $Y$. We already pointed out one such
effect, the underlying stellar absorption, which if not accounted for, 
results in the underestimation of $Y$. This effect is more pronounced for the
He {\sc i} $\lambda$4471 emission line. Recently, Gonz\'alez Delgado, Leitherer
\& Heckman (1999) have produced synthetic spectra of H Balmer and He {\sc i}
absorption lines in starburst and poststarburst galaxies. They predict that
the equivalent width of the He {\sc i} $\lambda$4471 absorption line for 
young starbursts with an age $t$ $\la$ 5 Myr, which is the case for 
Tol 1214--277 and Tol 65, can be in the range 0.4 -- 0.6 \AA. Comparing with 
equivalent widths of the He {\sc i} $\lambda$4471 emission line 7.6\AA\ and 
4.8\AA\ in Tol 1214--277 and Tol 65, we conclude that the underestimate of the
He abundance can be as high as $\sim$ 10\% in the case of the He {\sc i} 
$\lambda$4471 emission line. For the other two He {\sc i} lines 
$\lambda$5876 and
$\lambda$6678 lines the underestimate seems to be significantly smaller
because of larger emission line equivalent widths.
Unfortunately, Gonz\'alez Delgado et al. (1999) did not calculate equivalent
widths for the He {\sc i} $\lambda$5876 and $\lambda$6678 absorption lines. 
The weighted mean helium mass fraction in our calculations is mainly
defined by the strongest He {\sc i} $\lambda$5876 emission line with the 
highest weight. The upward correction of $Y$ for this line due to the 
underlying stellar absorption is not larger than $\sim$ 1\% for Tol 1214--277
and $\sim$ 1\% -- 2\% for Tol 65. Here we assume that the equivalent width
of the He {\sc i} $\lambda$5876 absorption line is 0.4\AA, which is $\sim$
100 times smaller than the equivalent width of the emission line in 
Tol 1214--277 and $\sim$ 80 times smaller in Tol 65 (Table \ref{Tab1}).

    Another source of systematic uncertainties comes from the assumption
that the H$^+$ and He$^+$ zones in the H {\sc ii} region are coincident. 
However, depending on the hardness of the ionizing radiation, the radius of the
He$^+$ zone can be smaller than the radius of the H$^+$ zone in
the case of soft ionizing radiation and larger in the case of hard
radiation. In the former case, a correction for unseen neutral helium
should be made, resulting in an ionization correction factor $ICF$(He) $>$ 1
and hence a higher helium abundance. In the latter case, the situation is 
opposite and $ICF$(He) $<$ 1. Furthermore, the electron temperature 
$T_{\rm e}$ in the O$^{+2}$ zone derived from the collisionally excited 
[O {\sc iii}] lines was assumed in our calculations to be constant and is the 
same as that in the H$^{+}$ and He$^{+}$ zones. However, 
$T_{\rm e}$(O {\sc iii}) tends to be larger than the temperature for 
the recombination lines of H {\sc i} and He {\sc ii} and, if applied, results
in an overestimate of the helium abundance. Both these effects have been 
discussed in several studies (e.g., Pagel et al. 1992; ITL97; Steigman, 
Viegas \& Gruenwald 1997; Olive et al. 1997; Viegas et al. 2000; 
Peimbert, Peimbert \& Ruiz 2000; Ballantyne et al. 2000; 
Sauer \& Jedamzik 2001). It was shown that the correction of the helium 
abundance can be as high as several percent in either downward or upward
directions depending on the hardness of the radiation. The hardness is
characterized by the ``radiation softness parameter'' $\eta$ defined
by V\'{\i}lchez \& Pagel (1988) as
\begin{equation}
\eta = \frac{{\rm O}^+}{{\rm S}^+}\frac{{\rm S}^{+2}}{{\rm O}^{+2}}.      
\label{eq:eta}
\end{equation}
Besides $\eta$ some other parameters have been used to derive
$ICF$(He), in particular the [O {\sc iii}] $\lambda$5007/H$\beta$ and 
[O {\sc iii}] $\lambda$5007/[O {\sc i}] $\lambda$6300 emission line flux
ratios (Ballantyne et al. 2000), the ionization parameter $U$, which is
the ratio of ionizing photon density to gas density, and the 
combination of all preceeding parameters (Sauer \& Jedamzik 2001).

    Sauer \& Jedamzik (2001) calculated an
extensive grid of photoionized H {\sc ii} region models aiming to derive the
correction factors as functions of $\eta$ and $U$. Their conclusion was that 
a downward correction of $Y$ as much as 6\% and 2\% is required respectively 
for ionization parameters log $U$ = --3.0 and --2.5 (see their Fig. 18).
However, the downward correction is $\la$ 1\% if log $U$ $\ga$ --2.0.

   For Tol 1214--277 and Tol 65 we find respectively log $\eta$ = --0.29 and 
--0.08. The ratios ([O {\sc iii}] $\lambda$4959 + 5007)/[O {\sc ii}] 
$\lambda$3727 of 19.9 and
7.2 in those galaxies (Table \ref{Tab1}) at an oxygen abundance 12 + log O/H =
7.54 correspond to an ionization parameter log $U$ $\ga$ --2.0 (McGaugh 1991).
In particular, Campbell (1988) derived log $U$ = --1.61$^{+0.10}_{-0.43}$ 
for Tol 1214--277 at the 70\% confidence level. With these $\eta$ and $U$
values, the downward correction of $Y$ due to ionization effects and variations
of
the electron temperature in Tol 1214--277 and Tol 65 is unlikely to be greater
than $\sim$ 1\%. Taking into account the fact 
that the upward correction of $Y$ due to
underlying stellar absorption can be as high as 1\% -- 2\%, we conclude
that both effects seem to offset each other and the combined 
systematic uncertainty
is $\la$ 1\% in Tol 1214--277 and Tol 65. Similar conclusions can be drawn
for SBS 0335--052 (Izotov et al. 1999) and SBS 0940+544 (Guseva et al. 2001).

Another approach has been developed by Peimbert (1967) to take into 
account the difference in the electron temperature in the O$^{++}$ zone 
as compared to the H$^{+}$ and He$^{+}$ zones. He developed a
formalism introducing an average temperature $T_0$ and a 
mean square temperature
variation $t^2$ in an H {\sc ii} region. Then the temperatures in the 
O$^{++}$ and H$^{+}$ and He$^{+}$ zones are expressed as different functions
of $T_0$ and $t^2$, and in hot H {\sc ii} regions $T_e$(O {\sc iii}) $\geq$ 
$T_e$(H {\sc ii}), $T_e$(He {\sc ii}). This approach has been applied 
by Peimbert et al. (2001) for the determination of the He 
abundance in some low-metallicity dwarf galaxies, including the two most-metal 
deficient BCDs, I Zw 18 and SBS 0335--052. They use the observations by
Izotov et al. (1999) and find that while the ionization correction
factors $ICF$(He) in both galaxies are very close to unity, the difference in 
$T_e$(O {\sc iii}) and $T_e$(He {\sc ii}) results in the reduction of the
He mass fraction by 2 -- 3 percent compared to the case with 
$T_e$(O {\sc iii}) = $T_e$(He {\sc ii}). Additionally, Peimbert et al. (2001) 
considered the effect of collisional excitation of hydrogen emission lines,
first noted by Davidson \& Kinman (1985). Neglecting this effect results in 
an artificially large extinction and hence overcorrection of the 
He {\sc i} $\lambda$5876 and $\lambda$6678 emission lines. Peimbert et al.
(2001) find that this effect in I Zw 18 and SBS 0335--052 leads to an the
upward correction of $Y$ by $\sim$ 2\%. Hence, after correction for all
the systematic effects considered, they obtain $Y$ = 0.241 and 0.245,
respectively
for I Zw 18 and SBS 0335--052. These values are similar to $Y$ = 0.243 and
0.246 derived earlier by Izotov et al. (1999) for those galaxies. 
The importance
of the correction for collisional excitation of the hydrogen emission
lines has been pointed out also by Stasi\'nska \& Izotov (2001) who concluded
that this effect can result in an upward $Y$ correction of up to 5\%, assuming
that the excess of the H$\alpha$/H$\beta$ flux ratio above the
theoretical recombination value is due only to collisional excitation.
However, in practice, some part of the
H$\alpha$/H$\beta$ flux ratio excess is due to interstellar extinction and
the correction of $Y$ for collisional excitation of the hydrogen lines is
likely smaller, only $\sim$ 2 -- 3 percent, similar to the value obtained
by Peimbert et al. (2001). Because the physical conditions in the H {\sc ii}
regions of SBS 0940+544 (Guseva et al. 2001), Tol 1214--277 and Tol 65 (this
paper) are similar to those in I Zw 18 and SBS 0335--052, we expect that
the estimates above for the systematic errors are valid for all the very 
metal-deficient high-excitation H {\sc ii} regions considered in this paper.

Besides the above effects, the uncertainties in the He {\sc i} 
recombination coefficients of $\sim$ 1.5\% also may play a role (Benjamin, 
Skillman \& Smits 1999). Although some systematic effects are still difficult
to estimate and are poorly studied, when taken into account 
together, it seems that they largely offset each other. Because of the 
uncertainties of these effects, we conservatively assume that combined 
systematic error in the He 
abundance determination may be tentatively set to 2\% (the error is 
2$\sigma$).

   Our studies show that the helium mass fraction in the 
lowest-metallicity BCDs observed with the Keck telescope lies
in the range 0.246 -- 0.247. Correction for the small contribution of $^4$He 
produced in stars $\Delta$$Y$ = 0.0010 -- 0.0017 (Izotov et al. 1999), 
results in a mean primordial $^4$He mass fraction $Y_{\rm p}$ 
of 0.245 $\pm$ 0.003 (rms) $\pm$ 0.005(sys) (2$\sigma$) 
obtained from the Keck 
observations of the four galaxies, in agreement with the
previous studies of ITL97, Izotov \& Thuan (1998) and Izotov et al. (1999). 
This $Y_{\rm p}$ predicts a baryon mass fraction $\Omega_{\rm b}h^2$ =
0.017$\pm$0.005(rms)$^{+0.010}_{-0.007}$(sys) (2$\sigma$), 
consistent with 0.020$\pm$0.002 (2$\sigma$)
derived from the primordial deuterium abundance (Burles \& Tytler 1998a, 1998b;
Burles et al. 2001). It is also consistent with the
estimation of high-$\ell$ peaks in the angular power spectrum of the
Cosmic Microwave Background (CMB) (Netterfield et al. 2001). This overall
consistency gives uniform support to the standard 
big bang nucleosynthesis model. In particular, if the baryon mass fraction
$\Omega_{\rm b}h^2$ = 0.022$\pm$0.003 (1$\sigma$) inferred from the CMB power 
spectrum is adopted, then, in the frame of the SBBN, the predicted 
primordial $^4$He mass fraction is $Y_{\rm p}$ = 0.248$\pm$0.001 (Lopez \&
Turner 1999; Burles, Nollett \& Turner 2001). Our $\Omega_{\rm b}h^2$
is also consistent with 
$\Omega_{\rm b}h^2$ = 0.025$\pm$0.001 (1$\sigma$) derived by Pettini 
\& Bowen (2001) from the deuterium abundance measurements in the 
$z_{\rm abs}$ = 2.0762 damped Lyman $\alpha$ system toward the QSO 2206--199
if 2$\sigma$ systematic error of 2\% in $Y_{\rm p}$ is assumed.

%
\section{Summary}
%
The main conclusions drawn from our Keck spectroscopic analysis of
the extremely metal-deficient BCDs Tol 1214--277 and Tol 65 may be summarized 
as follows:

1. The oxygen abundances in Tol 1214--277 and Tol 65 are 12 + log O/H =
7.54 $\pm$ 0.01, or 1/24 solar. We find that the nitrogen-to-oxygen
abundance ratio in both galaxies is log N/O = --1.64 $\pm$ 0.02, close
to the mean value of --1.60 found for the other most-metal deficient
BCDs with $Z$ $<$ $Z_\odot$/20 (Thuan et al. 1995; Izotov \& Thuan 1999). 
Alpha-product element-to-oxygen abundance ratios are in the same range
as those found for BCDs. The exception is the apparently higher Fe/O abundance
ratio in Tol 1214--277 which we argue is due to the contamination
of [Fe {\sc iii}] $\lambda$4658 emission line by C {\sc iv} $\lambda$4658
emission.

2. The $^4$He mass fractions in Tol 1214--277 and Tol 65 are respectively
$Y$ = 0.2466$\pm$0.0043 and 0.2463$\pm$0.0057. These values, after small
corrections for the helium produced in stars, correspond to a primordial
$^4$He mass fraction of 0.245, in excellent agreement with previous
studies of ITL97, Izotov \& Thuan (1998) and Izotov et al. (1999), supporting
the validity of the standard big bang nucleosynthesis model.

3. We find that the systematic uncertainties in the $^4$He abundance
determination in Tol 1214--277 and Tol 65 due to the combined effect
of underlying stellar absorption, temperature and ionization structure 
of the H {\sc ii} region and collisional excitation of the hydrogen emission 
lines are likely small, not exceeding $\sim$ 2\% (2$\sigma$).


\acknowledgements
Y.I.I. thanks for partial financial support through NATO grant PST.EV.976026,
Swiss SCOPE grant 7UKRJ62178 and for hospitality
at National Optical Astronomical Observatory. 

\clearpage


\clearpage

%
%

\begin{deluxetable}{lccrcccr}
\tablenum{1}
\tablecolumns{8}
\tablewidth{0pt}
\tablecaption{Emission line fluxes and equivalent widths \label{Tab1}}
\tablehead{ &\multicolumn{3}{c}{Tol 1214--277}&&\multicolumn{3}{c}{Tol 65} \\
\cline{2-4} \cline{6-8}
\colhead{Ion}&\multicolumn{1}{c}{$F$($\lambda$)/$F$(H$\beta$)}&\multicolumn{1}{c}{$I$($\lambda$)/$I$(H$\beta$)}&
\colhead{$EW$}&&\multicolumn{1}{c}{$F$($\lambda$)/$F$(H$\beta$)}&\multicolumn{1}{c}{$I$($\lambda$)/$I$(H$\beta$)}
&\colhead{$EW$}}
\startdata
 3727\ [O {\sc ii}]              &0.332$\pm$0.006&0.341$\pm$0.006&   42.3&&0.634$\pm$0.010&0.674$\pm$0.011&   67.1\\
 3750\ H12                       &0.031$\pm$0.002&0.033$\pm$0.003&    4.1&&0.012$\pm$0.001&0.046$\pm$0.007&    1.3\\
 3770\ H11                       &0.034$\pm$0.002&0.035$\pm$0.003&    4.5&&0.024$\pm$0.002&0.059$\pm$0.005&    2.7\\
 3798\ H10                       &0.050$\pm$0.002&0.052$\pm$0.003&    6.8&&0.033$\pm$0.002&0.068$\pm$0.004&    3.6\\
 3835\ H9                        &0.052$\pm$0.002&0.054$\pm$0.003&    7.1&&0.046$\pm$0.002&0.082$\pm$0.004&    4.9\\
 3868\ [Ne {\sc iii}]            &0.343$\pm$0.006&0.351$\pm$0.006&   47.3&&0.246$\pm$0.004&0.259$\pm$0.005&   26.1\\
 3889\ He {\sc i} + H8           &0.203$\pm$0.004&0.208$\pm$0.004&   28.2&&0.154$\pm$0.003&0.197$\pm$0.004&   16.2\\
 3968\ [Ne {\sc iii}] + H7       &0.291$\pm$0.005&0.298$\pm$0.005&   43.3&&0.208$\pm$0.004&0.251$\pm$0.005&   21.9\\
 4026\ He {\sc i}                &0.019$\pm$0.001&0.019$\pm$0.001&    3.0&&0.012$\pm$0.002&0.013$\pm$0.002&    1.3\\
 4069\ [S {\sc ii}]              &0.008$\pm$0.001&0.008$\pm$0.001&    1.2&&0.011$\pm$0.002&0.011$\pm$0.002&    1.2\\
 4101\ H$\delta$                 &0.268$\pm$0.005&0.274$\pm$0.005&   45.6&&0.222$\pm$0.004&0.260$\pm$0.005&   25.8\\
 4227\ [Fe {\sc v}]              &0.007$\pm$0.001&0.007$\pm$0.001&    1.3&&\nodata        &\nodata        &\nodata\\
 4340\ H$\gamma$                 &0.481$\pm$0.008&0.488$\pm$0.008&   96.9&&0.443$\pm$0.007&0.477$\pm$0.008&   58.0\\
 4363\ [O {\sc iii}]             &0.167$\pm$0.003&0.169$\pm$0.003&   34.0&&0.093$\pm$0.002&0.095$\pm$0.002&   12.3\\
 4471\ He {\sc i}                &0.035$\pm$0.001&0.035$\pm$0.001&    7.6&&0.034$\pm$0.002&0.034$\pm$0.002&    4.8\\
 4658\ [Fe {\sc iii}]            &0.005$\pm$0.001&0.005$\pm$0.001&    1.2&&0.006$\pm$0.001&0.006$\pm$0.001&    1.0\\
 4686\ He {\sc ii}               &0.049$\pm$0.002&0.049$\pm$0.002&   12.1&&0.012$\pm$0.001&0.012$\pm$0.001&    1.9\\
 4711\ [Ar {\sc iv}] + He {\sc i}&0.027$\pm$0.001&0.027$\pm$0.001&    6.7&&0.011$\pm$0.001&0.011$\pm$0.001&    1.8\\
 4740\ [Ar {\sc iv}]             &0.016$\pm$0.001&0.016$\pm$0.001&    4.0&&0.006$\pm$0.001&0.006$\pm$0.001&    0.9\\
 4861\ H$\beta$                  &1.000$\pm$0.015&1.000$\pm$0.015&  267.5&&1.000$\pm$0.015&1.000$\pm$0.015&  174.1\\
 4922\ He {\sc i}                &0.012$\pm$0.001&0.012$\pm$0.001&    3.2&&0.011$\pm$0.001&0.010$\pm$0.001&    1.9\\
 4959\ [O {\sc iii}]             &1.707$\pm$0.025&1.703$\pm$0.025&  478.3&&1.243$\pm$0.019&1.212$\pm$0.018&  225.7\\
 5007\ [O {\sc iii}]             &5.100$\pm$0.075&5.082$\pm$0.075& 1467.2&&3.733$\pm$0.055&3.628$\pm$0.055&  693.4\\
 5200\ [N {\sc i}]               &\nodata        &\nodata        &\nodata&&0.007$\pm$0.001&0.007$\pm$0.001&    1.4\\
 5876\ He {\sc i}                &0.093$\pm$0.002&0.091$\pm$0.002&   41.4&&0.111$\pm$0.002&0.103$\pm$0.002&   31.4\\
 6300\ [O {\sc i}]               &0.012$\pm$0.001&0.011$\pm$0.001&    6.1&&0.020$\pm$0.001&0.019$\pm$0.001&    6.7\\
 6312\ [S {\sc iii}]             &0.008$\pm$0.001&0.008$\pm$0.001&    4.2&&0.011$\pm$0.001&0.010$\pm$0.001&    3.8\\
 6363\ [O {\sc i}]               &0.003$\pm$0.001&0.003$\pm$0.001&    1.5&&0.008$\pm$0.001&0.007$\pm$0.001&    2.6\\
 6563\ H$\alpha$                 &2.822$\pm$0.042&2.737$\pm$0.044& 1571.1&&3.074$\pm$0.045&2.769$\pm$0.045& 1079.6\\
 6583\ [N {\sc ii}]              &\nodata        &\nodata        &\nodata&&0.020$\pm$0.001&0.018$\pm$0.001&    4.7\\
 6678\ He {\sc i}                &0.026$\pm$0.001&0.025$\pm$0.001&   15.4&&0.031$\pm$0.001&0.027$\pm$0.001&   11.2\\
 6717\ [S {\sc ii}]              &0.024$\pm$0.001&0.023$\pm$0.001&   14.1&&0.071$\pm$0.002&0.064$\pm$0.002&   26.7\\
 6731\ [S {\sc ii}]              &0.021$\pm$0.001&0.020$\pm$0.001&   12.7&&0.052$\pm$0.001&0.046$\pm$0.001&   19.7\\
 7065\ He {\sc i}                &0.025$\pm$0.001&0.024$\pm$0.001&   17.1&&0.031$\pm$0.001&0.028$\pm$0.001&   13.6\\
 7135\ [Ar {\sc iii}]            &0.022$\pm$0.001&0.022$\pm$0.001&   15.6&&0.035$\pm$0.001&0.031$\pm$0.001&   15.7\\
 7281\ He {\sc i}                &0.006$\pm$0.001&0.005$\pm$0.001&    4.0&&0.006$\pm$0.001&0.005$\pm$0.001&    2.9\\
 7320\ [O {\sc ii}]              &0.005$\pm$0.001&0.005$\pm$0.001&    3.8&&0.016$\pm$0.001&0.014$\pm$0.001&    7.6\\
 7330\ [O {\sc ii}]              &0.004$\pm$0.001&0.004$\pm$0.001&    3.0&&0.011$\pm$0.001&0.010$\pm$0.001&    5.4\\ \\
 $C$(H$\beta$) dex    &\multicolumn {3}{c}{0.040$\pm$0.019} &&\multicolumn {3}{c}{0.115$\pm$0.019}\\
 $F$(H$\beta$)\tablenotemark{a} &\multicolumn{3}{c}{ 1.88$\pm$0.01}&&\multicolumn{3}{c}{ 2.65$\pm$0.01}\\
 $EW$(abs)\ \AA      &\multicolumn{3}{c}{0.1}&&\multicolumn{3}{c}{3.5}\\
\enddata
\tablenotetext{a}{in units of 10$^{-14}$ erg\ s$^{-1}$cm$^{-2}$.}
\end{deluxetable}

\clearpage

\begin{deluxetable}{lccc}
\tablenum{2}
\tablecolumns{4}
\tablewidth{0pt}
\tablecaption{Heavy element abundances \label{Tab2}}
\tablehead{
\colhead{Parameter}&\colhead{Tol 1214--277}&&\colhead{Tol 65}}
\startdata
$T_{\rm e}$(O {\sc iii})(K)                     &19790$\pm$260  &&17320$\pm$240 \\
$T_{\rm e}$(O {\sc ii})(K)                      &15630$\pm$190  &&14770$\pm$200 \\
$T_{\rm e}$(S {\sc iii})(K)                     &18130$\pm$210  &&16080$\pm$200 \\
$N_{\rm e}$(S {\sc ii})(cm$^{-3}$)              &   400$\pm$120 &&   50$\pm$50 \\ \\
O$^+$/H$^+$($\times$10$^5$)                     &0.273$\pm$0.010&&0.614$\pm$0.024\\
O$^{+2}$/H$^+$($\times$10$^5$)                  &2.982$\pm$0.095&&2.816$\pm$0.101\\
O$^{+3}$/H$^+$($\times$10$^5$)                  &0.191$\pm$0.011&&0.046$\pm$0.005\\
O/H($\times$10$^5$)                             &3.447$\pm$0.096&&3.477$\pm$0.104\\
12 + log(O/H)                                   &7.538$\pm$0.012&&7.541$\pm$0.013\\ \\
N$^{+}$/H$^+$($\times$10$^7$)                   &\nodata        &&1.402$\pm$0.065\\
ICF(N)\tablenotemark{a}                         &\nodata        &&5.66\,~~~~~~~~~~\\
log(N/O)                                        &\nodata        &&--1.642$\pm$0.024~~\\ \\
Ne$^{+2}$/H$^+$($\times$10$^5$)                 &0.418$\pm$0.014&&0.425$\pm$0.016\\
ICF(Ne)\tablenotemark{a}                        &1.16\,~~~~~~~~~~&&1.23\,~~~~~~~~~~\\
log(Ne/O)                                       &--0.854$\pm$0.019~~&&--0.821$\pm$0.021~~\\ \\
S$^+$/H$^+$($\times$10$^7$)                     &0.414$\pm$0.014&&1.118$\pm$0.031\\
S$^{+2}$/H$^+$($\times$10$^7$)                  &2.304$\pm$0.228&&4.272$\pm$0.353\\
ICF(S)\tablenotemark{a}                         &2.90\,~~~~~~~~~~&&1.69\,~~~~~~~~~~\\
log(S/O)                                        &--1.640$\pm$0.039~~&&--1.581$\pm$0.031~~\\ \\
Ar$^{+2}$/H$^+$($\times$10$^7$)                 &0.576$\pm$0.024&&0.981$\pm$0.036\\
Ar$^{+3}$/H$^+$($\times$10$^7$)                 &1.252$\pm$0.087&&0.596$\pm$0.129\\
ICF(Ar)\tablenotemark{a}                        &1.01\,~~~~~~~~~~&&1.03\,~~~~~~~~~~\\
log(Ar/O)                                       &--2.273$\pm$0.025~~&&--2.332$\pm$0.039~~\\ \\
Fe$^{+2}$/H$^+$($\times$10$^7$)                 &0.864$\pm$0.222\tablenotemark{b}&&1.187$\pm$0.274\\
ICF(Fe)\tablenotemark{a}                        &15.8\,~~~~~~~~~~&&7.07\,~~~~~~~~~~\\
log(Fe/O)                                       &--1.404$\pm$0.112\tablenotemark{b}~~&&--1.617$\pm$0.101~~\\
$[$O/Fe$]$                                      &--0.017$\pm$0.112~~&&0.197$\pm$0.101\\
\enddata
\tablenotetext{a}{ICF is the ionization correction factor for unseen
stages of ionization. The expressions for ICFs are adopted from ITL94.}
\tablenotetext{b}{Probable overestimate; see text.}
\end{deluxetable}

\clearpage

\begin{deluxetable}{lccc}
\tablenum{3}
\tablecolumns{4}
\tablewidth{0pt}
\tablecaption{Helium abundance \label{Tab3}}
\tablehead{
\colhead{Parameter}&\colhead{Tol 1214--277}&&\colhead{Tol 65}}
\startdata
$T_{\rm e}$(O {\sc iii})(K)                     &19790$\pm$260  &&17320$\pm$240 \\
$N_{\rm e}$(He {\sc ii})(cm$^{-3}$)             &    25$\pm$1   &&  150$\pm$50 \\
$\tau$($\lambda$3889)                           &      0.01     &&      0.01   \\ \\
$y^+$($\lambda$4471)                &0.0755$\pm$0.0029&&0.0706$\pm$0.0033\\
$y^+$($\lambda$5876)                &0.0773$\pm$0.0016&&0.0812$\pm$0.0022\\
$y^+$($\lambda$6678)                &0.0773$\pm$0.0028&&0.0796$\pm$0.0031\\
$y^+$(weighted mean)                &0.0770$\pm$0.0013&&0.0784$\pm$0.0016\\
$y^+$($\lambda$5876 + $\lambda$6678)&0.0773$\pm$0.0014&&0.0807$\pm$0.0018\\
$y^{+2}$($\lambda$4686)             &0.0046$\pm$0.0001&&0.0011$\pm$0.0001\\ \\
$y$($\lambda$4471)                        &0.0801$\pm$0.0029&&0.0717$\pm$0.0033\\
$y$($\lambda$5876)                        &0.0819$\pm$0.0016&&0.0823$\pm$0.0022\\
$y$($\lambda$6678)                        &0.0819$\pm$0.0028&&0.0796$\pm$0.0031\\
$y$(weighted mean)                        &0.0816$\pm$0.0013&&0.0795$\pm$0.0016\\
$y$($\lambda$5876 + $\lambda$6678)        &0.0819$\pm$0.0014&&0.0818$\pm$0.0018\\ \\
$Y$($\lambda$4471)                         &0.2424$\pm$0.0090&&0.2226$\pm$0.0107\\
$Y$($\lambda$5876)                         &0.2466$\pm$0.0050&&0.2475$\pm$0.0069\\
$Y$($\lambda$6678)                         &0.2466$\pm$0.0087&&0.2439$\pm$0.0098\\
$Y$ (weighted mean)                        &0.2458$\pm$0.0039&&0.2410$\pm$0.0050\\
$Y$ ($\lambda$5876 + $\lambda$6678)        &0.2466$\pm$0.0043&&0.2463$\pm$0.0057\\
\enddata
\end{deluxetable}

\clearpage

\clearpage

\clearpage

\end{document}